\documentstyle[preprint,aps]{revtex}
\tighten

 \input amssym.def   
 \input amssym.tex

 \input BoxedEPS
\SetRokickiEPSFSpecial  
 \HideDisplacementBoxes

\font\twelvemsb=msbm10 at 12pt
\font\ninemsb=msbm7 at 9pt
\font\sixmsb=msbm5 at 6pt
\newfam\Msbfam
\textfont\Msbfam=\twelvemsb
\scriptfont\Msbfam=\ninemsb
\scriptscriptfont\Msbfam=\sixmsb
\def\Bbb#1{{\fam\Msbfam\relax#1}}

\def\half{{\textstyle{1\over2}}}
\def\quar{{\textstyle{1\over4}}}
\def\eigh{{\textstyle{1\over 8}}}

\def\onestwo{{\textstyle{1\over\sqrt{2}}}}
\def\beq{\begin{equation}}
\def\eeq{\end{equation}}
\def\bi{\begin{itemize}}
\def\ei{\end{itemize}}
\def\beqar{\begin{eqnarray}}
\def\eeqar{\end{eqnarray}}

\newcommand{\Ee}{\mbox{$\cal E\;$}}

\newcommand{\Pp}{\mbox{$\cal P\;$}}
\newcommand{\bcP}{\mbox{\boldmath$\cal P$}}

\def\vec#1{{\bf#1}}
\newcommand{\pa}{\partial}

\newcommand{\rmd}{{\rm d\null}}
\newcommand{\rmdd}[1]{\rmd^d#1\,}
\def\boldnab{\mbox{\boldmath$\nabla$}}
\def\boldr{{\bf r}}
\def\pmbb#1{\mbox{\footnotesize\boldmath$#1$}}
\def\pmb#1{\mbox{\boldmath$#1$}}

\let\varkappa\kappa

\skip\footins = 14pt 
\begin{document}


\title{Nonlinear Realization of a Dynamical Poincar\'e Symmetry\\
               by a Field-dependent Diffeomorphism}

\author{D. Bazeia\footnotemark[1] and R. Jackiw\footnotemark[2]}

\footnotetext[1] {\baselineskip=12pt On leave from Departamento de
F\'\i sica, Universidade Federal da Para\'\i ba, Caixa Postal 5008,
58051-970 Jo\~ao Pessoa, Para\'\i ba, Brazil. Partially supported by
Conselho Nacional de Desenvolvimento Cient\'\i fico e Tecnol\'ogico,
CNPq, Brazil.}

\footnotetext[2] {\baselineskip=12pt This work is supported in part by funds
provided by  the U.S.~Department of Energy (D.O.E.) under contract
\#DE-FC02-94ER40818.}

\address{Center for Theoretical Physics\\ Massachusetts Institute of
Technology\\ Cambridge, MA ~02139--4307}

\maketitle
\begin{abstract}\noindent
We consider a description of membranes by (2, 1)-dimensional field theory,
or alternatively a description of irrotational, isentropic fluid motion
by a field theory in any dimension. We show that these
Galileo-invariant systems, as well as others related to them,
admit a peculiar diffeomorphism symmetry, where the transformation
rule for coordinates involves the fields. The symmetry algebra
coincides with that of the Poincar\'e group in one higher
dimension. Therefore, these models provide a nonlinear representation
for a dynamical Poincar\'e group.
\end{abstract}

\vskip 1cm

\begin{center}
Submitted to {\it Annals of Physics}\\
MIT CTP\# 2717 
\end{center}



\section{Introduction}

\indent
When a dynamical system is invariant under some transformation,
typically the dynamical variables transform in a naturally evident,
linear fashion. However, it may happen that in special circunstances,
there exist unexpected invariances,
frequently called ``dynamical symmetries,'' and the relevant
transformation law for the dynamical variables is intricate and nonlinear.
An example is given by the motion of a particle in the $1/r$ potential. There
is obvious invariance against the $O(3)$ group of rotation
transformations, which rotate the particle position variable ${\vec r}$.
Additionally, there is a hidden $O(3,1)$ or $O(4)$ symmetry, which acts in an
unexpected manner on the dynamical variables.

In this paper we shall present a field theoretic instance of this phenomenon.
We shall consider a family of field Lagrangians, which arise in various
physical contexts. We shall show that these theories possess a hidden
Poincar\'e invariance, where the Poincar\'e transformations are defined in
one higher dimension, and act nonlinearly on the dynamical variables of the
model.

\section{Dynamical Model}
\indent
Consider the following field Lagrangian, defined on a $d$-dimensional space
$\{ \boldr\}$, and describing first-order evolution  in time $\{ t \}$
\beq
L= \int \rmdd r \Bigr( \theta \dot{\rho} - \half \rho {\boldnab} \theta \cdot 
{\boldnab} \theta - V(\rho)\Bigl)
\label{eq:1}
\eeq
The fields depend on time and space: $\rho(t, \boldr)$, $\theta(t, \boldr)$;
an over-dot always indicates differentiation with respect to the time argument,
the gradient is always with respect to spatial arguments.  $V$ is an
unspecified $\theta$-independent potential that provides interactions for
the free Lagrangian
\beq
L_0= \int \rmdd r \Bigr( \theta \dot{\rho} -
\half \rho {\boldnab} \theta \cdot 
{\boldnab}\theta \Bigl)
\label{eq:2}
\eeq
The symplectic structure indicates that $\theta$ and $\rho$ are canonically
conjugate\cite{ref:1}. The Euler--Lagrange equations read
\begin{mathletters}\label{eq:12}
\begin{eqnarray}
\dot{\rho} &=& - \boldnab \cdot (\rho \boldnab \theta)
\label{eq:12a}\\
\dot{\theta}&=& -\half (\boldnab \theta)^2 + f(\rho)
\label{eq:12b}\\
f(\rho)&\equiv&-\frac{\delta}{\delta \rho} \int \rmdd r V
\label{eq:12c}
\end{eqnarray}
\end{mathletters}
where the ``force'' term $f(\rho)$ is absent in the free case.

One encounters the dynamics described by $L_0$ and $L$ in diverse branches of
physics:  fluid motion with no vorticity and pressure determined by $V$
(irrotational and isentropic motion)\cite{ref:2}, quantum mechanics in a
hydrodynamical formulation\cite{ref:3,ref:4}, membrane theory\cite{ref:5},
and dimensional reduction of relativistic scalar field theory\cite{ref:6}.
Here is one instructive derivation of $L_0$\cite{ref:4}.

In the free particle Lagrangian (mass set to unity)
\beq
L_{\rm free~particle}= \half \sum_i v_i^2 (t)
\label{eq:3}
\eeq
we replace the discrete summation with a continuum integration and
introduce a density $\rho$ and a current ${\bf j}$, 
\beq
{\bf j} (t, \boldr )= {\bf v} (t, \boldr) \rho (t, \boldr)
\label{eq:4}
\eeq
linked by a continuity equation
\beq
\dot{\rho} + \boldnab \cdot {\bf j} = 0\ \ .
\label{eq:5}
\eeq
Evidently (\ref{eq:3}) becomes $\half \int \rmdd r v^2 \rho =
\half \int \rmdd r j^2/\rho$, and  (\ref{eq:5}) is implemented by a Lagrange
multiplier $\theta$.
\beq
L'_0 = \int \rmdd r \Bigl\{ \half j^2/\rho +
\theta (\dot{\rho} + \boldnab \cdot {\bf j})
\Bigr\}
\label{eq:6}
\eeq
Since the symplectic term does not contain ${\bf j}$, it may be
eliminated by solving for it\cite{ref:1}.
\beq
{\bf j} = \rho \boldnab \theta
\label{eq:7}
\eeq
Eqs.~(\ref{eq:4}) and (\ref{eq:7}) exhibit the irrotational character of the
velocity ($\boldnab\times \vec v = 0$) whose potential is $\theta$
($\vec v = \boldnab\theta$), while substitution of (\ref{eq:7}) into
(\ref{eq:6}) gives $L_0$ of (\ref{eq:2}). $V$, whose form in (\ref{eq:1})
is arbitrary, provides an isentropic pressure potential for the
irrotational fluid motion described by $L_0$.

To see more cleary the connection with the equations of isentropic fluid
mechanics, when the velocity field ${\vec v}$ is irrotational
$({\boldnab}\times \vec v=0)$, we note first that Eq.~$(\ref{eq:12a})$ is just
the continuity equation. Moreover by taking the gradient of $(\ref{eq:12b})$,
we find
\beq
\dot{\vec v}+\vec v \cdot{\boldnab}{\vec v}=f'(\rho)\,{\boldnab}\rho
\eeq
This is Euler's equation, provided we identify
$f'(\rho)\,{\boldnab}\rho=-(1/\rho)\,{\boldnab}(pressure)$. But such an
identification is always possible for isentropic motion (Kelvin's theorem),
where presssure is a function of $\rho$ only. In that case
$(\partial V/\partial\rho)=-f(\rho)$ coincides with the enthalpy, and
$[\rho(\partial^2 V/\partial\rho^2)]^{1/2}$ is the sound speed.

For an alternative derivation of the Lagrangian $(\ref{eq:1})$ from a
different framework, we begin with the (linear or nonlinear)
Schr\"odinger theory Lagrangian
\beq
L_{\rm S} = \int \rmdd r \Bigl\{ i \psi^* \dot{\psi} -
\half (\boldnab\psi^*)\cdot(\boldnab \psi) -
\bar{V} (\psi^*\psi) \Bigr\}
\label{eq:8}
\eeq
with $\bar{V}$ determining any nonlinear interactions.
The substitution
\beq
\psi = \rho^{1/2} e^{i\theta}
\label{eq:9}
\eeq
produces $L$ of (\ref{eq:1})  (apart from a total time derivative) with $V$
fixed at 
\beq
V (\rho) = \bar{V} (\rho) + \eigh \frac{(\boldnab \rho)^2}{\rho}
\label{eq:10}
\eeq
This is the hydrodynamical form of the Schr\"odinger
theory\cite{ref:3,ref:4}: the kinetic term is $L_0$, but there is also a
further nontrivial ``interaction,'' even in the absence of $\bar{V}$.
The same result is obtained from a dimensional reduction of a
scalar relativistic field theory\cite{ref:6}.  

Finally, we recall that a gauge-fixed formulation of a membrane in $(3,1)$
Minkowski space-time\cite{ref:5} again leads to (\ref{eq:1}), with $d=2$,
and a specific potential of strength~$g$
\beq
V (\rho) = \frac{g}{\rho}
\label{eq:11}
\eeq
[One obtains the same result for a ``d-brane'' moving in $(d+1,1)$ space-time.]

\section{Obvious Symmetries of the Model}
\indent
From its derivation, it is clear that (\ref{eq:1})
and (\ref{eq:2}) (with obvious restriction on $V$) possess the Galileo
symmetry.  For completeness, we list here the generators of the infinitesimal
transformations, as integrals of the appropriate densities; also we specify
the action of the finite transformation (parameterized by
$\omega$) on the fields: $\rho\to \rho_\omega, \theta\to \theta_\omega$,
by presenting formulas for $\rho_\omega (t, \boldr)$ and
$\theta_\omega (t, \boldr)$ in terms of $\rho(t, \boldr)$ and $\theta(t,
\boldr)$.  One verifies that the generators are time independent according to
the equations of motion $(\ref{eq:12})$ and this furthermore implies that
the transformed fields $\rho_\omega$ and $\theta_\omega$ also solve
$(\ref{eq:12})$.
\newpage
\begin{itemize}
\item Time, space translation
   \begin{itemize}
   \item  Energy 
\beq
H = \int \rmdd r \Ee \ , \quad \Ee = \half \rho \boldnab \theta \cdot
\boldnab \theta + V(\rho) = \half j^2/\rho + V(\rho)
\label{eq:14}
\eeq

\item Momentum 
\beq
{\bf P} = \int \rmdd r \bcP \ , \quad
 \bcP =
\rho
\boldnab
\theta = {\bf j}
\label{eq:15}
\eeq
   \end{itemize}


\item Space rotation

    \begin{itemize}
    \item Angular momentum 
\beq
J^{ij} = \int \rmdd r(r^i \Pp^j - r^j \Pp^i)
\label{eq:16}
\eeq
    \end{itemize}
\end{itemize}
With these space-time transformations $\rho_\omega$, $\theta_\omega$ are
obtained from $\rho$, $\theta$ by respectively translating the time,
space arguments and by rotating the spatial argument.

\begin{mathletters}\label{eq:17}
\begin{itemize}
\item  Galileo boost
       \bi 
       \item Boost generator
          \beq
          {\bf B} = t {\bf P} - \int \rmdd r \boldr \rho
          \label{eq:17a}
          \eeq
       \ei
\end{itemize}
The boosted fields are
\begin{eqnarray}
\rho_{\pmbb\omega}^{\phantom{\prime}} (t, \boldr) &=& \rho (t, \boldr -
{\pmb \omega} t)
\label{eq:17b}
\\
\theta_{\pmbb\omega}^{\phantom{\prime}} (t,\boldr)&=&\theta (t, \boldr -{\pmb
\omega} t)+{\pmb \omega}\cdot \vec r -{\pmb \omega}^2t/2 \label{eq:17c}
\end{eqnarray}
The inhomogeneous terms in $\theta_{\pmbb\omega}^{\phantom{\prime}} $ are
recognized as the well-known Galileo 1-cocycle, compare (\ref{eq:9}).
Also they ensure that the transformation law for $\vec v = \boldnab \theta$
\beq
\vec v(t, \vec r) \to \vec v_{\pmbb \omega}^{\phantom{\prime}}(t,\vec r) =
\vec v(t, \vec r -\pmb \omega t) +\pmb\omega
\label{eq:17d}
\eeq
\end{mathletters}
is appropriate for a co-moving velocity.
Furthermore, knowledge about the Galileo 2-cocycle leads us to examine the
${\bf P},{\bf B}$ bracket, and its extension exposes another conserved
generator, arising from an invariance against translating $\theta$ by a
constant; this just reflects the phase arbitrariness in (\ref{eq:9}).
\begin{mathletters}\label{eq:18}
\begin{itemize}
\item Phase symmetry
     \bi
     \item Charge
\begin{eqnarray}
 N &=& \int \rmdd r \rho \label{eq:18a} \\
\rho_\omega &=& \rho \label{eq:18b}  \\
\theta_\omega  &=& \theta - \omega \label{eq:18c}
\end{eqnarray}
     \ei
\end{itemize}
\end{mathletters}

\section{Connection with Poincar\'e symmetry}
\indent
It is well known that a Poincar\'e group in $(d+1, 1)$ dimensions possesses
the above extended Galileo group as a subgroup.\cite{ref:7} This is seen by
identifying selected light-cone components of the Poincar\'e generators
${\Bbb P}^\mu, {\Bbb M}^{\mu\nu}$ with the Galileo generators,
\begin{eqnarray}
{\Bbb P}^\mu &=& ({\Bbb P}^-,{\Bbb P}^+, {\Bbb P}^i) = (H,N, P^i)
\label{eq:19} \\
{\Bbb M}^{\mu\nu} &=& ({\Bbb M}^{+-}, {\Bbb M}^{-i}, {\Bbb M}^{+i},
{\Bbb M}^{ij})
\nonumber \\
{\Bbb M}^{+i} &=& B^i, \quad {\Bbb M}^{ij} = J^{ij}  
\label{eq:20} 
\end{eqnarray}
where the $\pm$ components of tensors are defined by
\beq
T^{(\pm)} = \onestwo (T^{(0)} \pm T^{(d+1)}) \ \ .
\label{eq:21} 
\eeq
But the Lorentz generators ${\Bbb M}^{+-}$ and ${\Bbb M}^{-i}$
have no Galilean counterparts.

A remarkable fact, first observed in Ref.~\cite{ref:5} and then again in
simplified form in Ref.~\cite{ref:6}, is that in the model (\ref{eq:1})
with $V(\rho) = g/\rho$ [as in the membrane application, eq.~(\ref{eq:11})]
one can define quantities that can be set equal to the generators missing
from the identification  (\ref{eq:20}), namely $ {\Bbb M}^{+-}$
and ${\Bbb M}^{-i}$.  This holds for arbitrary interaction strength $g$;
setting it to zero allows the same construction for the free Lagrangian
(\ref{eq:2}). We wish to determine what role the additional generators
have for the models  (\ref{eq:1}) and  (\ref{eq:2}).

\section{Additional Symmetries}
\indent
We observe that the free action $I_0=\int \rmd t\, L_0$, as well as the
interacting one
\beq
I_g = \int \rmd t\,\rmdd r (\theta \dot{\rho} - \half \rho \boldnab
\theta \cdot \boldnab
 \theta - g/\rho) 
\label{eq:22} 
\eeq
are invariant against time rescaling $t \to e^{\omega} t$,
which is generated by
\begin{mathletters}\label{eq:23}
\begin{eqnarray}
D &=& tH - \int \rmdd r \rho \theta\ \ .   \label{eq:23a} \\
\noalign{\hbox{Fields  transform  according  to}}
\rho (t, \boldr) &\to& \rho_\omega (t, \boldr) = e^{-\omega} \rho
(e^\omega t, \boldr) \label{eq:23b} 
\\
\theta (t, \boldr) &\to& \theta_\omega (t, \boldr) =  e^{\omega} \theta
(e^\omega t, \boldr)\ \ .  \label{eq:23c}
\end{eqnarray}
\end{mathletters}
The dilation generator $D$ is identified with ${\Bbb M}^{+-}$.
It is straightforward to
verify from (\ref{eq:12}) that $D$ is indeed time independent.

More intricate is a further, obscure symmetry whose generator can be
identified with ${\Bbb M}^{-i}$.  Consider the field-dependent coordinate
tranformations, implicitly defined by
\begin{eqnarray}
t&\to& T (t,\boldr)=t+\half {\pmb \omega} \cdot (\boldr +{\bf R} (t, \boldr))
\nonumber
\\
\boldr &\to& {\bf R} (t, \boldr) =\boldr + {\pmb \omega} \theta(T, {\bf R}) 
\label{eq:24}
\end{eqnarray}
with Jacobian $\left| J \right|$
\beq
J = \det 
\left(
\begin{array}{rcl}
\displaystyle \frac{\partial T}{\partial t} && \displaystyle \frac{\partial
T}{\partial r^j}
\\[2ex]
\displaystyle \frac{\partial R^i}{\partial t} && \displaystyle
 \frac{\partial R^i}{\partial r^j} 
\end{array}
\right) = 
\Bigl( 1 - {\pmb \omega} \cdot \boldnab \theta (T,{\bf R}) - \half \omega^2
\dot{\theta} (T, {\bf R}) \Bigr)^{-1}
\label{eq:25}
\eeq
The transformation parameter ${\pmb \omega}$ has dimensions of inverse
velocity. When fields are taken to transform according to 
\begin{mathletters}
\begin{eqnarray}
\rho (t, \boldr)&\to& \rho_{\pmbb\omega}^{\phantom{\prime}} (t, \boldr)=
\rho (T, {\bf R})
\frac{1}{|J|}
\label{eq:26a}
\\
\theta (t, \boldr) &\to& \theta_{\pmbb\omega}^{\phantom{\prime}} (t, \boldr)=
\theta (T, {\bf R})
\label{eq:26b}
\end{eqnarray}\label{eq:26}
\end{mathletters}
one verifies that $I_g$ and $I_0$ are invariant.  This is readily seen for the
interaction term
\beq
g \int \rmd t\, \rmdd r\, \frac1{\rho(t, \boldr)} \to g \int \rmd t\,\rmdd r\,
\frac{|J|}{\rho(T, {\bf R})} =
g\int \rmd T\, \rmdd R  \frac1{\rho(T, {\bf R})}\ \ .
\label{eq:27}
\eeq
To establish invariance of $I_0$, it is useful to write it first as
\beqar
I_0 &=& -\int \rmd t\,\rmdd r\,\rho(\dot\theta + \half \boldnab\theta\cdot
\boldnab\theta)\nonumber\\
&& \to  -\int \rmd t\,\rmdd r\, \frac{\rho(T,\vec R)}{|J|} \Bigl\{
\frac\pa{\pa t} \theta(T,\vec R) + \half \frac\pa{\pa r^i}\theta(T,\vec R) 
\frac\pa{\pa r^i}\theta(T,  \vec R) \Bigr\}
\label{eq:28}
\eeqar
The desired result follows once it is realized that the quantity in curly
brackets equals $J^2\bigl\{ \dot\theta(T,\vec R) +
\half \left(\boldnab\theta(T,\vec R)\right)^2\bigr\}$. The
transformations (\ref{eq:26}) are generated by 
\beqar
\vec G &=& \int \rmdd r \left(\vec r \Ee - \half \rho \boldnab\theta^2
 \right) \nonumber\\
 &=& \int \rmdd r \left(\vec r \Ee - \theta\pmb{\cal P}
 \right)
\label{eq:29}
\eeqar
which is time independent according to (\ref{eq:12}).

While we have no insight about the geometric aspects of this peculiar
symmetry, the following remarks may help achieve some transparency. 

Observe that  the Galileo generators can be expressed in terms of $\rho$
and $\vec j$ or $\rho$ and $\vec v=\vec j/\rho = \boldnab \theta$.
Consequently, they are also defined for velocity fields with vorticity,
($\boldnab\times \vec v \neq 0$), and provide well-known constants of motion
for the (isentropic) Euler equations\cite{ref:2}. However, the velocity
potential~$\theta$ is needed to form
$D$ and $\vec G$, which therefore have a role only in vortex-free motion
(with a specific potential or no potential).

For gauge-fixed membrane theory in (3, 1)-dimensional space-time,
the dynamical variables
are the membrane's transverse coordinates $r^i$, ($i=1,2$), which are
functions of time~$t$
and of two parametric variables $\phi^i$, ($i=1,2$). The quantity
$1/\rho$ arises as the Jacobian  for the transformation $(t,\pmb {\bf\phi})
\to (t, \pmb r(t,\pmb {\bf\phi}))$\cite{ref:5}. Therefore it is natural
that under the further transformation $(t, \vec r)\to\bigl( T(t, \vec r),
\vec R(t, \vec r)\bigr)$, $1/\rho$ acquires the Jacobian of that
transformation. Presumably transformations (\ref{eq:24}), (\ref{eq:26})
reflect a residual invariance of gauge-fixed membrane theory, but the reason
for the specific form (\ref{eq:24}) of the transformation is not apparent.
From the identification with Poincar\'e generators, we see that $\rho$ is
the ${\Bbb P}^+$ density and $\theta$ is in some sense like~$x^-$, and
indeed in membrane theory $\theta$ coincides with
$x^-(t,\pmb {\bf\phi})\to x^-(t, \vec r)$.

Note that for all the transformations that are identified with Lorentz
transformations, {\it viz}. $(\ref{eq:16})$, $(\ref{eq:17})$, $(\ref{eq:23})$
and $(\ref{eq:24})-(\ref{eq:26})$, the following relation holds between new
(capitalized) and old (lower-case) coordinates
\beq
2\,T\,\theta(T, \vec R)- R^2 = 2\,t\,\theta_{\pmbb\omega}(t,\vec r)- r^2
\eeq
The naturalness of this relation is recognized once it is appreciated that
in light-cone components $x^{\mu}x_{\mu}=2x^+x^- -x^i x^i$.

In applications to fluid mechanics our transformation generates nontrivial
solutions of Euler's equations, as we now explain.

\section{Transforming Explicit Solutions}
\indent
In order to gain insight into the peculiar diffeomorphism
transformations~(\ref{eq:26}), we consider its effect on some explicit
solutions to eqs.~(\ref{eq:12}), in the free ($V=0$) and
interacting ($V=g/\rho$) cases.

\subsection{No interaction, \protect\boldmath$V\, \mbox{\bf= 0}$}
\indent
With $V=0$, eq.~(\ref{eq:12b}) is solved by
\begin{mathletters}
\beq
 \theta (t, \vec r) = \frac12 \frac{r^2}t
 \label{eq:30}
\eeq
which, apart from selecting an origin in time and space  and presenting a
rotation and boost invariant profile, is also invariant against time
rescaling~(\ref{eq:23}) and the unconventional
diffeomorphism (\ref{eq:24})--(\ref{eq:26}). The fluid moves with a velocity
unaffected by boosts~(\ref{eq:17d})
\beq
\vec v = \frac{\vec r}t\ \ .
\label{eq:31}
\eeq
The density is not determined, since the solution of the continuity
equation (\ref{eq:12a}) in $d$~spatial dimensions involves an arbitrary
function of $t/r$, and of the angles specifying~$\vec r$
\beq
 \rho(t,\vec r) = \frac{f(t/r, \hat\vec  r)}{r^d}\ \ .
\label{eq:32}
\eeq
\end{mathletters}
With $\theta$ as in (\ref{eq:30}), the coordinate
transformations~(\ref{eq:24}) take the explicit forms
\beqar
T(t,\vec r) &=& t   r_{\pmbb\omega}^2  \nonumber\\
\vec R(t,\vec r) &=& r \vec r_{\pmbb\omega} \nonumber\\
 J &=& r_{\pmbb\omega}^2 \nonumber\\
\vec r_{\pmbb\omega} &\equiv& \hat\vec  r + \frac{\pmb\omega}2 \frac rt
\label{eq:33}
\eeqar
so that, as stated, the transformed $\theta_{\pmbb\omega}^{\phantom{\prime}}
(t, \boldr)$ coincides with $\theta (t, \boldr)$, while the transformed
density becomes a different function of $t/r$,
\beq
 \rho_{\pmbb\omega}(t,\vec r)  ={\displaystyle
\frac{f\Bigl({\displaystyle\frac{t 
r_{\pmbb\omega}}r , 
  {\displaystyle   \hat\vec  r_{\pmbb\omega}}\Bigr)}}{ r^d\,  
r_{\pmbb\omega}^{d+2}}}\ \ .
\label{eq:34}
\eeq
This coincides with $\rho(t,\vec r)$ for the special choice $ f(t/r,
\hat\vec  r) \propto   ( t/r)^{2+d}$   in~(\ref{eq:32}), which provides a
density profile that is invariant under the diffeomorphism 
(\ref{eq:24})--(\ref{eq:26}).

One may construct other ``free'' solutions, for which~$\theta$ remains
unchanged under (\ref{eq:24}), (\ref{eq:25}), (\ref{eq:26b}), while~$\rho$
involves arbitrary functions. Also there are free solutions that respond
nontrivially to the transformations. We do not pursue
any of this any further here, rather we examine the ``interacting'' 
problem.

\subsection{With interaction, \protect\boldmath$V\, \mbox{\bf=}\, g/\rho$}
\indent
Remarkably,  a solution of the form (\ref{eq:30}) also works in the
presence of interactions.
The profiles
\begin{mathletters}
\beqar
\theta(t,\vec r) &=& -\frac{r^2}{2(d-1)t} \label{eq:35a}\\
\rho(t,\vec r) &=& \sqrt{\frac{2g}{d}} (d-1) \frac{\left|t\right|}r
\label{eq:35b}
\eeqar
solve (\ref{eq:12}) with $V=g/\rho$; evidently $d$ must be greater than $1$,
and $g$ positive. [The positivity requeriment is natural, in view of the
fact that the sound speed for our model is
$[\rho(\partial^2 V/\partial\rho^2)]^{1/2}=\sqrt{2g}/\rho$.]
This solution is time-rescaling invariant, but no longer boost nor
diffeomorphism invariant. The velocity flow is given by 
\beq
\vec v = -\frac{\vec r}{(d-1)t} 
\label{eq:36}
\eeq
while the current reads
\beq
\vec j = \mp \sqrt{\frac{2g}d}\, \hat{\vec r}\ \ .
\label{eq:36d}
\eeq
where the sign is determined by the sign of~$t$. 
\end{mathletters}

At $d=1$ we can find time-rescaling invariant solutions for $g>0$ 
 \beqar
 \theta(t, x) &=& \frac1{2k^2 t} \sinh^2kx, \mkern22mu -\frac1{2k^2
t}\cosh^2kx\nonumber\\
 \rho(t,x) &=&  \frac{\sqrt{2g}\,k\left|t\right|}{\sinh^2 kx}, \mkern68mu 
\frac{\sqrt{2g}\, k\left|t\right|}{\cosh^2 kx}
\label{eq:37}
 \eeqar
 while for $g<0$ one gets
  \beqar
 \theta(t,x) &=& \frac1{2k^2 t} \sin^2kx, \mkern34mu \frac1{2k^2t}\cos^2kx
\nonumber\\
 \rho(t,x) &=& \frac{\sqrt{2\left|g\right|}\,k\left|t\right|}{\sin^2 kx},
\mkern42mu
 \frac{\sqrt{2\left|g\right|}\,k\left|t\right|}{\cos^2 kx}\ \ .
 \label{eq:38}
  \eeqar
 Here $k$ is an arbitrary, positive integration constant. (The above are
general solutions that preserve the scale of time, since a second
integration constant is the origin of~$x$.)
 
 We now exhibit the form of the solutions when the field-dependent, coordinate
transformations (\ref{eq:24})--(\ref{eq:26}) are carried out. For
simplicity we discuss only the
$d=2$ (membrane) case, and take $t>0$. The new coordinates are determined
by the old ones by (\ref{eq:24}), (\ref{eq:25}), and (\ref{eq:35a}). 
\begin{mathletters}
 \beqar
 T&=& {\textstyle\frac34} t + \half{\pmb\omega}\cdot\vec r \pm\quar
 \sqrt{(t+2{\pmb\omega}\cdot\vec r)^2 - 2\omega^2 r^2} \label{eq:39a}\\
 \vec R &=& \vec r + \frac{\pmb\omega}{2\omega^2} 
 \Bigl[ -t - 2{\pmb\omega}\cdot\vec r \pm 
\sqrt{(t+2{\pmb\omega}\cdot\vec r)^2 -
2\omega^2 r^2} \Bigr]\label{eq:39b}\\
\frac1J&=& 1 + \frac{{\pmb\omega}\cdot\vec R}T - \frac{\omega^2 R^2}{4T^2}
\label{eq:39c}
 \eeqar
\end{mathletters}
After $\theta$ and $\rho$ are transformed according to the rules
(\ref{eq:26}), it is noticed that expressions are simplified by performing
the Galileo boost $\vec r \to \vec r - {\pmb\omega}t/\omega^2$,
according to the rules~(\ref{eq:17}). (This precludes taking the
limit ${\pmb\omega}\to0$.) Also, time is rescaled according to
(\ref{eq:23}), with $t\to\sqrt2 t$. Finally, we define ${\pmb\omega}/\omega^2
= \vec c$, which has dimension of velocity, and then the transformed
profiles provide two solutions, depending on the sign of the square root 
\begin{mathletters}\label{eq:40}
\beqar
\theta_{\vec c} (t,\vec r) &=& \pm \sqrt{2(\vec c \cdot \vec r)^2 - c^2r^2-
c^4 t^2}
\label{eq:40a}\\
 \rho_{\vec c} (t,\vec r) &=& \frac{\sqrt{2g}}{c^2} \biggl[\frac{{2(\vec c
\cdot \vec r)^2 - c^2r^2 - c^4 t^2}}{ r^2 \mp 2t\sqrt{2(\vec c \cdot
\vec r)^2 - c^2r^2 - c^4 t^2}  }\biggr]^{1/2}~.
\label{eq:40b}\\
\noalign{\hbox{The velocity is}}
\vec v_{\vec c}(t,\vec r) &=& \pm \frac{2\vec c( \vec c\cdot\vec r) -
\vec r c^2}{\sqrt{2(\vec c \cdot \vec r)^2 - c^2r^2 - c^4 t^2}}
\label{eq:40c}\\
\noalign{\hbox{and the current reads}}
\vec j_c(t,\vec r) &=& \pm \sqrt{2g} 
\frac{2\hat\vec c (\hat\vec c\cdot\vec r)-\vec r}{\Bigl[r^2 \mp 2t
        \sqrt{2(\vec c\cdot \vec r)^2 - c^2r^2 - c^4 t^2}\Bigr]^{1/2}} \ \ .
\label{eq:40d}
\eeqar
\end{mathletters}
Note that $c$ may be replaced by $ic$, and $\rho_c$ by $-\rho_c$,
to obtain another solution.

In the Figures we exhibit the profiles of the interacting solutions.
We plot the  original and
transformed densities, and the transformed currents
$\vec j_c = \rho_c\boldnab \theta_c$, in terms of the variables
$\vec r/t$ ($t>0$). Without loss of generality, $\vec c$ is taken
along the $x$-axis, and its magnitude is incorporated in the dimensionless
ratio $\vec r/ct$.
The original density possesses a singularity at the origin; in the
transformed solutions the
singularity is present only with the upper (negative) sign in the bracketed
expression of (\ref{eq:40b}), where its denominator vanishes at $r^2 =
(\hat\vec c\cdot\vec r)^2 = 2c^2t^2$.
The transformed currents exhibit a similar singularity. In the physical
region the argument of the square root must be positive,
$2(\hat\vec c\cdot \vec r)^2 - r^2 - c^2 t^2 \ge 0$.

\newpage

\begin{figure}
$$\BoxedEPSF{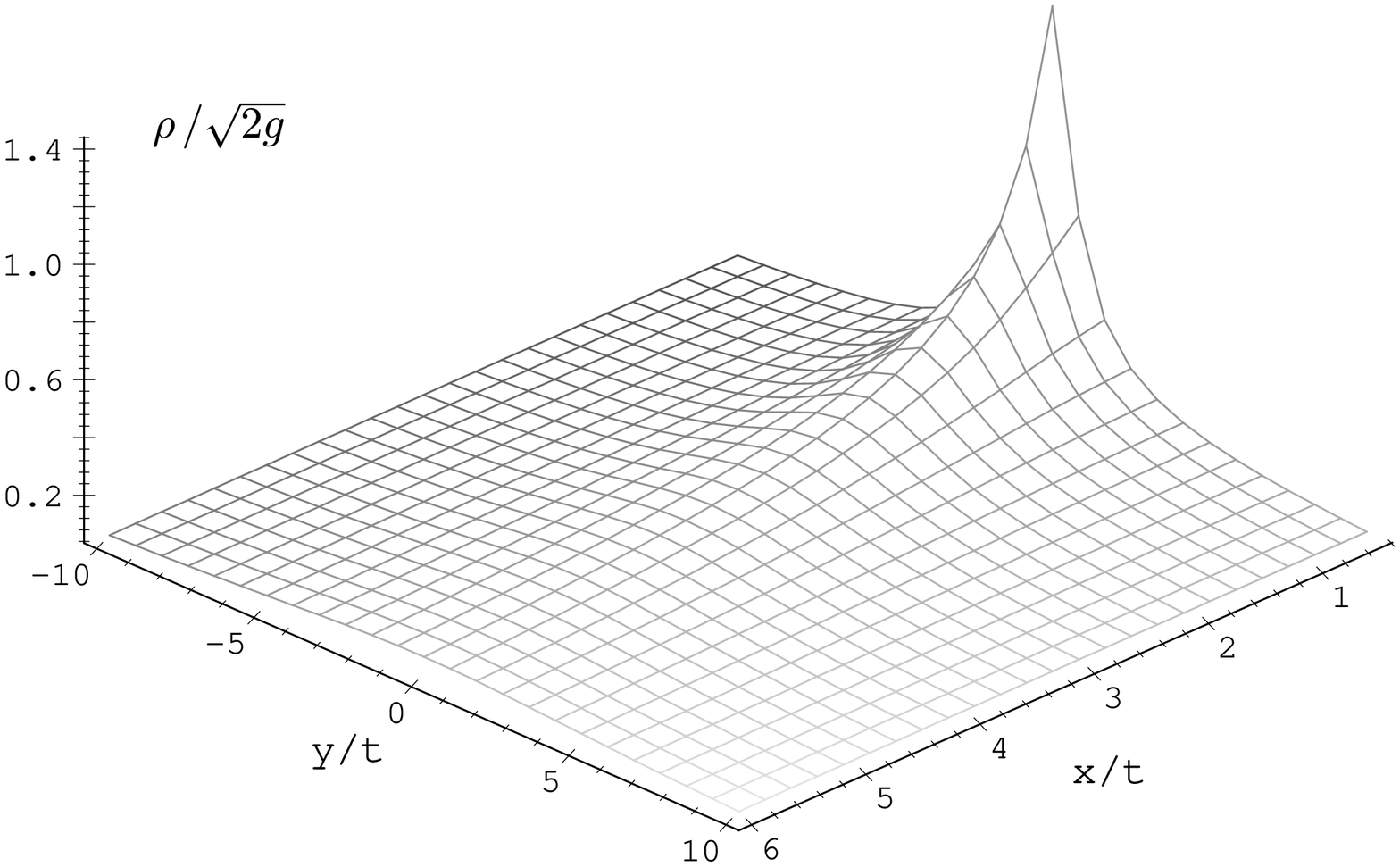 scaled 700}$$
\caption{The original density  $\rho(t,\vec r)/\protect{\sqrt{2  g}}$.}
\label{DBRJfig:1}
\end{figure}

\begin{figure}
$$\BoxedEPSF{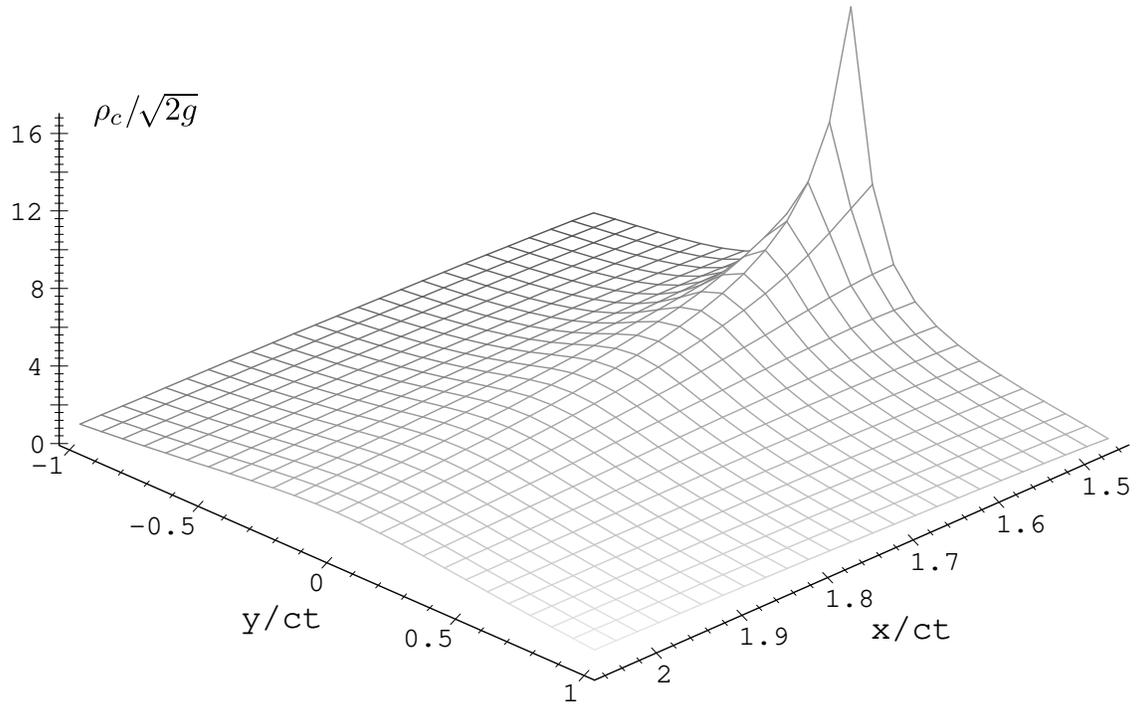 scaled 700}$$
\caption{The transformed density  $\rho_c(t,\vec r)/\protect{\sqrt{2  g}}$,
with the upper sign.}
\label{DBRJfig:2}
\end{figure}

\begin{figure}
$$\BoxedEPSF{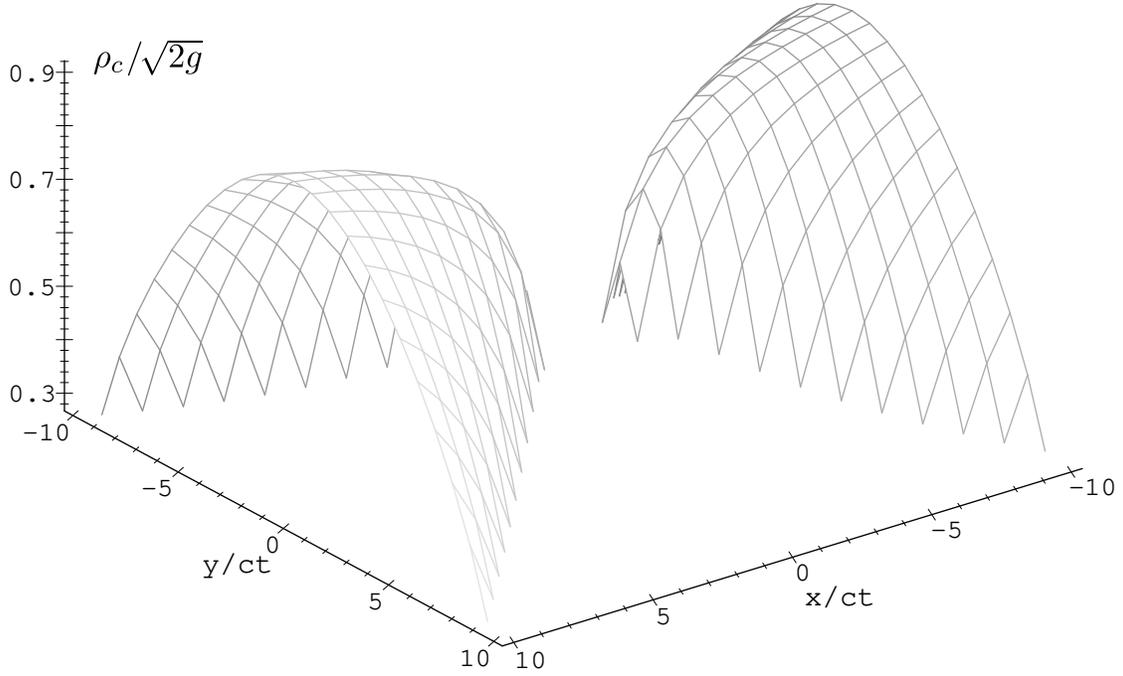 scaled 666}$$
\caption{The transformed density  $\rho_c(t,\vec r)/\protect{\sqrt{2 g}}$,
with the lower sign. The envelope defining the physical region
is at $x^2-y^2=c^2t^2$.}
\label{DBRJfig:3}
\end{figure}

\begin{figure}
$$\BoxedEPSF{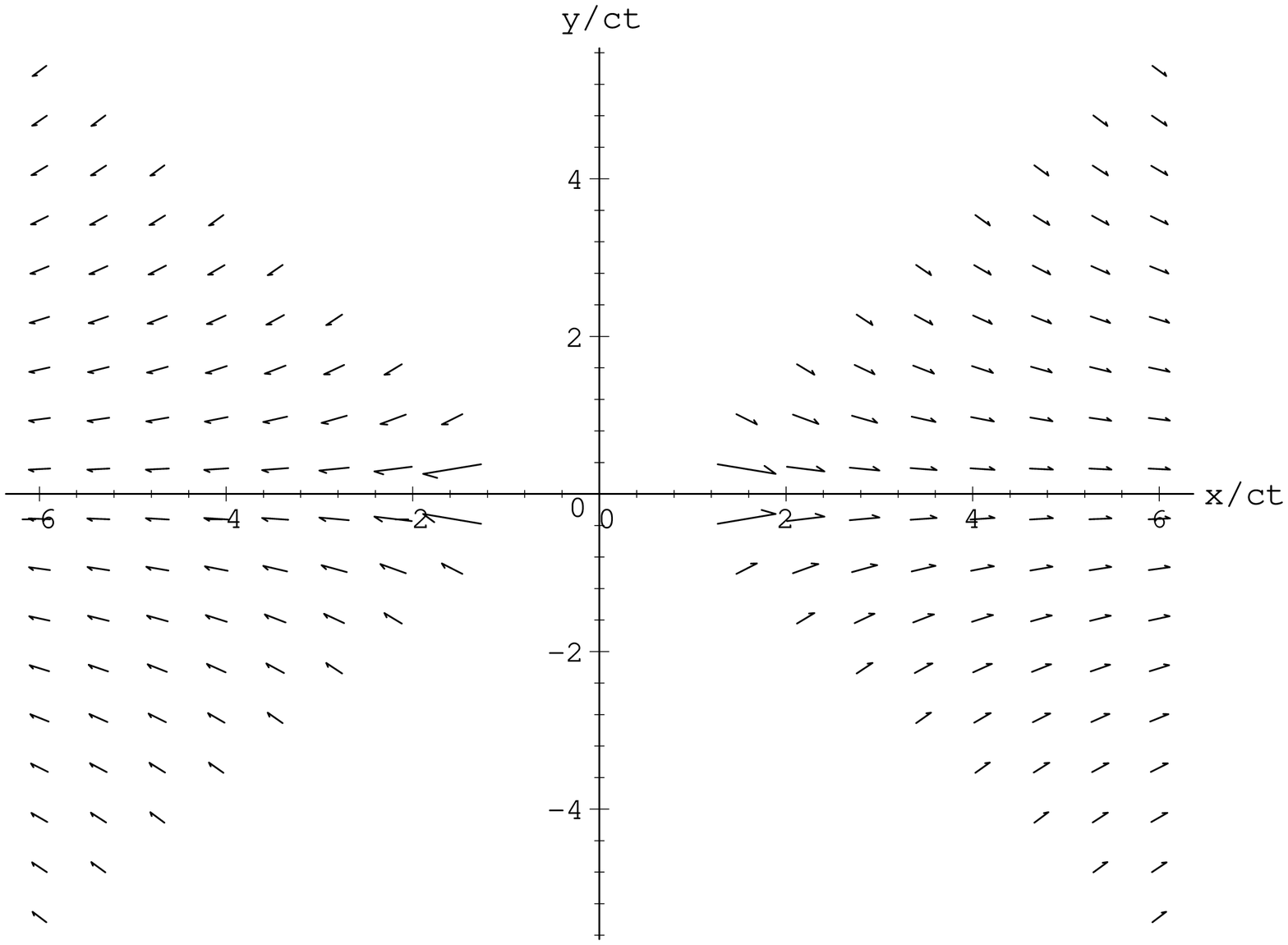 scaled 600}$$
\caption{The transformed current  $\vec j_c(t, \vec r)/\protect{\sqrt{2 g}}$,
with the upper signs.  The envelope defining the physical region
is at $x^2-y^2=c^2t^2$.}
\label{DBRJfig:4}
\end{figure}

\begin{figure}
$$\BoxedEPSF{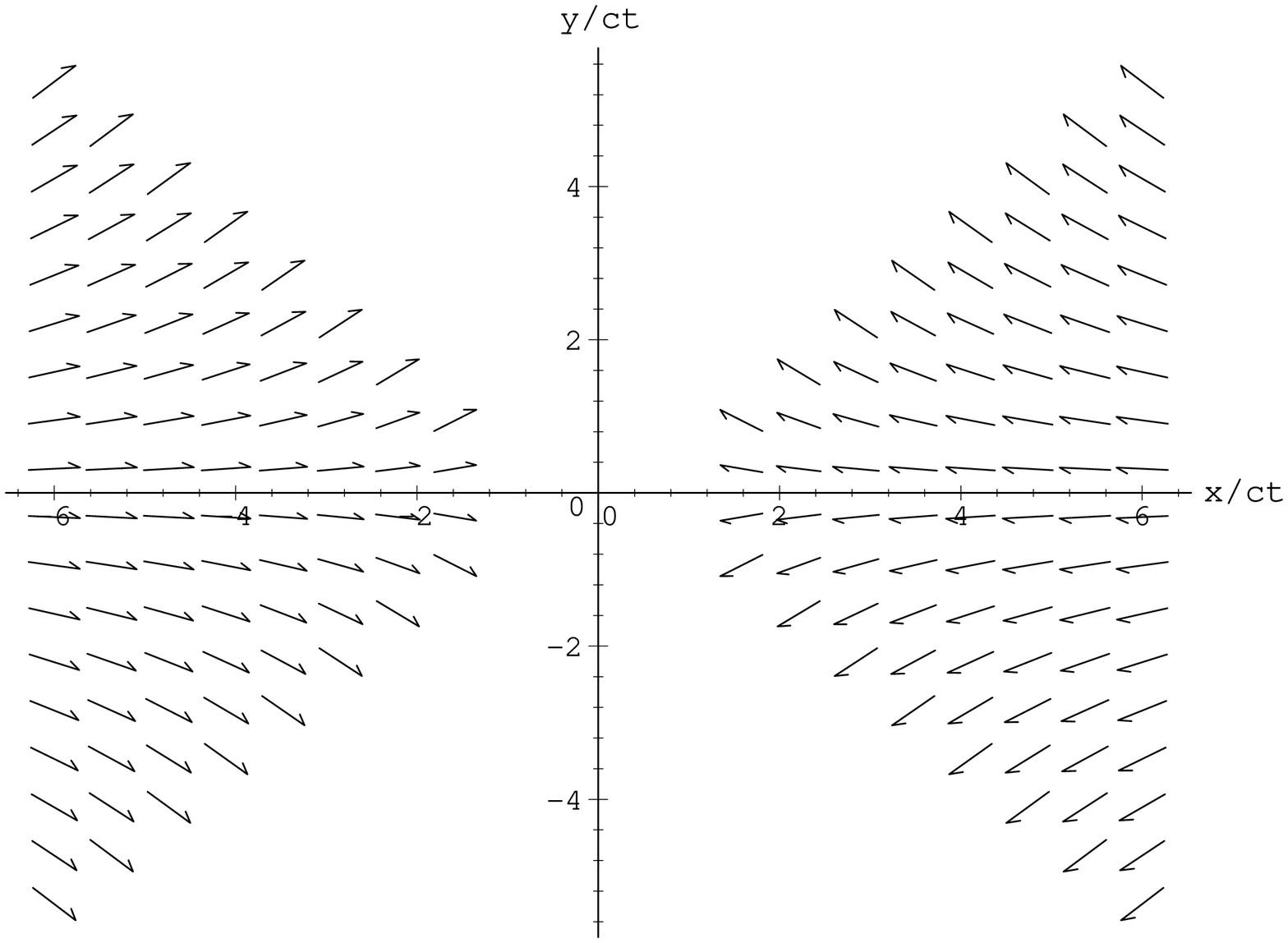 scaled 600}$$
\caption{The transformed current $\vec j_c(t,
\vec  r)/\protect{\sqrt{2  g}}$, with the lower signs. The envelope
defining the physical region is at $x^2-y^2=c^2t^2$.}
\label{DBRJfig:5}
\end{figure}


\end{document}